# Studying RNA homology and conservation with Infernal: from single sequences to RNA families

**Lars Barquist[1,2], Sarah W. Burge[2], Paul P. Gardner[3,4]**

[1]Institute for Molecular Infection Biology, University of Würzburg, Würzburg, D-97080 Germany

[2]Wellcome Trust Sanger Institute, Hinxton, Cambridge, CB10 1SA United Kingdom; Fax: +44 (0)1223 494919

[3]School of Biological Sciences, University of Canterbury, Private Bag 4800, Christchurch, New Zealand

[4]Biomolecular Interaction Centre, University of Canterbury, Private Bag 4800, Christchurch, New Zealand

Correspondence to: Lars Barquist (lars.barquist@uni-wuerzburg.de)

## Significance Statement

Determining the phylogenetic distribution of ncRNAs is challenging. Unlike protein sequences, many ncRNA sequences are relatively short and lack obvious patterns in their primary sequence that can be used for de novo prediction and homology detection. As a result, specialized computational and curatorial techniques, discussed in detail here, are required to perform accurate homology search and annotation transfer when starting with a single known ncRNA sequence. This protocol will provide readers with the background necessary to begin defining their own ncRNA families suitable for use in comparative, functional, and evolutionary studies of structured RNA elements.

## Abstract

Emerging high-throughput technologies have led to a deluge of putative non-coding RNA (ncRNA) sequences identified in a wide variety of organisms. Systematic characterization of these transcripts will be a tremendous challenge. Homology detection is critical to making maximal use of functional information gathered about ncRNAs: identifying homologous sequence allows us to transfer information gathered in one organism to another quickly and with a high degree of confidence. ncRNA presents a challenge for homology detection, as the primary sequence is often poorly conserved and de novo secondary structure prediction and search remains difficult. This protocol introduces methods developed by the Rfam database for identifying "families" of homologous ncRNAs starting from single "seed" sequences using manually curated sequence alignments to build powerful statistical models of sequence and structure conservation known as covariance models (CMs), implemented in the Infernal software package. We provide a step-by-step iterative protocol for identifying ncRNA homologs, then constructing an alignment and corresponding CM. We also work through an example for the bacterial small RNA MicA, discovering a previously unreported family of divergent MicA homologs in genus *Xenorhabdus* in the process.

## Introduction

Over the past twenty years, the development and commodification of high-throughput technologies for reading DNA has led to dramatic changes in the way biology is done. The development of reliable and inexpensive transcriptomic technologies, in particular, has lead to major changes to our understanding of the role RNA plays in both bacterial and eukaryotic systems (Barquist and Vogel, 2015; Rinn and Chang, 2012). While exceptions to the central dogma have been known since the discovery of tRNA in the late 1950's, it was only with the development of genome-scale sequencing technologies that the pervasive nature of regulation mediated by non-coding RNA (ncRNA) has become clear. This has led to a diverse menagerie of RNA classes, ranging from the microRNAs (miRNAs) and bacterial small RNAs (sRNAs) that operate to modulate gene expression through RNA:RNA antisense interactions, to small nucleolar RNAs (snoRNAs) that guide RNA modification enzymes, to the riboswitches capable of sensing metabolites with so-called RNA aptamers, to the diverse long non-coding RNAs (lncRNAs) suspected to blanket eukaryotic genomes. Characterizing a single ncRNA is a challenge in itself; to maximize the utility of this work, it is often desirable to transfer functional annotations between organisms. For instance, one might identify an ncRNA associated with disease in humans, but want to study it in a system where technical and (more importantly) ethical considerations allow for genetic manipulation, such as mice. Similarly in bacteria, molecular tools are often best developed in model strains such as *E. coli*, making them attractive platforms for characterizing ncRNAs, which can then be inferred to operate in a similar function in related bacteria. However, this inference depends crucially on computational tools capable of transferring these hard-won functional annotations through homology prediction.

This problem of homology prediction is usually approached as the problem of finding sequences which align well to our source sequence, on the reasonable assumption that similar sequences are more likely to be evolutionarily and functionally related. A wide range of critical applications in genomics rely on our ability to produce "good" alignments. Single-sequence homology search as implemented in tools such as BLAST (Altschul et al., 1990) is an (often heuristic) application of alignment. The sensitivity and specificity of homology search can be improved by the use of evolutionary information in the form of accurate substitution and insertion-deletion (indel) rates derived from multiple sequence alignments (MSAs), captured in the statistical models used by HMMER (Finn et al., 2011, 2015; Eddy, 2011) and Infernal (Nawrocki et al., 2009; Nawrocki and Eddy, 2013) for protein and RNA alignments respectively. These models can be thought of as defining "families" of homologous sequences, as in the Pfam and Rfam databases (Finn et al., 2014; Nawrocki et al., 2015). By using these models to classify sequences, we can infer functional and structural properties of uncharacterized sequences.

Unfortunately, producing the high-quality "seed" alignments of RNA these methods require remains difficult. While proteins can be aligned accurately using only primary sequence information with pairwise sequence identities as low as 20% for an average-length sequence (Rost, 1999; Thompson et al., 1999), it appears that the "twilight zone" where blatantly erroneous alignments occur between RNA sequences may begin at above 60% identity (Gardner et al., 2005) (Gardner et al., 2005)(Lindgreen et al., 2014). The inclusion of secondary structure information can improve alignment accuracy (Freyhult et al., 2007), but predicting secondary structure is not trivial (Gardner and Giegerich, 2004; Puton et al., 2014). An instructive example of the difficulties this can lead to is the case of the 6S gene, a bacterial ncRNA which modulates $\sigma^{70}$ activity during the shift from exponential to stationary growth. The *Escherichia coli* 6S sequence was determined in 1971 (Brownlee, 1971) and its function determined in 2000 (Wassarman and Storz, 2000). However, the extent of this gene's phylogenic distribution was not realized until 2005 when Barrick and colleagues carefully constructed an alignment from a number of deeply diverged putative 6S sequences, and through successive secondary-structure aware homology searches demonstrated its presence across large swaths of the bacterial phylogeny (Barrick et al., 2005). Even now, new homologs are discovered on a regular basis (Sharma et al., 2010; Weinberg et al., 2010; Wehner et al., 2014), and 6S appears to be an ancient and important component of the bacterial regulatory machinery. Similar examples of the power of enhanced homology search considering both sequence and structural information can be found for other classes of ncRNAs, such as riboswitches (Barrick and Breaker, 2007), ribosomal leader sequences (Fu et al., 2013), ribozymes (Weinberg et al., 2015), and snoRNAs (Gardner et al., 2010).

In this chapter, it is our hope to make these techniques accessible to sequence analysis novices. We introduce the techniques necessary to construct a high-quality RNA alignment from a single seed sequence, and then use the information contained in this alignment to identify additional more distant homologs, expanding the alignment in an iterative fashion. These methods, while time-consuming, can be far more sensitive than a BLAST search (Menzel et al., 2009). We present a brief protocol which starts with a single sequence, and then uses a collection of web and command-line based tools for alignment, structure prediction, and search to construct an Infernal covariance model (CM), a probabilistic model which captures many important features of structured RNA sequence variation (Nawrocki and Eddy, 2013). These models may then be used in the iterative expansion of alignments or for homology search and genome annotation. CMs are also are used by the Rfam database in defining RNA sequence families, and are the subject of a dedicated RNA families track at the journal **RNA Biology** (Gardner and Bateman, 2009). We include as an instructive example the construction of an RNA family for the enterobacterial small RNA MicA, used as the basis for the current Rfam model, discovering a convincing divergent clade of homologs in the process.

## Strategic Planning

A large variety of tools exist for RNA sequence analysis. Given the diversity of ncRNA sequence, structure, and function, it is likely no one set of tools will work optimally for every ncRNA. Here we review a number of alternative methods and tools that can be easily substituted at various stages of our protocol.

### Single Sequence Search

We rely on NCBI BLAST (Altschul et al., 1990) to quickly identify close homologs of RNA sequences in this protocol. NCBI and EMBL-EBI both maintain servers (Boratyn et al., 2013; McWilliam et al., 2013) with slightly different interfaces, though there are no substantive differences in the implementations. We use the NCBI server here. EBI also maintains servers for a number of BLAST and FASTA derivatives, which may be helpful. Both sites also allow users to BLAST against databases of expressed sequences including GEO at NCBI, and high throughput cDNA and transcriptome shotgun assembly databases at EMBL-EBI (Barrett et al., 2013; Silvester et al., 2015). Such searches can be helpful for gathering comparative expression data for your ncRNA.

A nucleotide version of the HMMER3 package (Eddy, 2011) for profile-based sequence search provides both increased sensitivity and specificity over BLAST at little additional computational cost (Wheeler and Eddy, 2013). We hope that a web server similar to the one currently available for protein sequences (Finn et al., 2015) will be forthcoming. In the meantime, RNAcentral (RNAcentral Consortium, 2015; Bateman et al., 2011) offers an nhmmer-based search facility, however it is limited to searching known ncRNA sequences. If it is possible that homologous sequences are spliced (e.g. introns in the U3 snoRNA (Myslinski et al., 1990)), then a splice-site aware search method may be useful, such as BLAT (Kent, 2002) or GenomeWise (Birney et al., 2004), but there are not publicly available webservers for them that we are aware of.

| **Resource** | **Reference** | **URL** |
|---|---|---|
| NCBI-BLAST | (Johnson et al., 2008) | http://blast.ncbi.nlm.nih.gov/Blast.cgi |
| EMBL-EBI NCBI-BLAST | (McWilliam et al., 2013) | http://www.ebi.ac.uk/Tools/sss/ncbiblast/ |
| EMBL-EBI Sequence Search | (McWilliam et al., 2013) | http://www.ebi.ac.uk/Tools/sss/ |
| HMMER3 | (Finn et al., 2015, 2011) | http://www.ebi.ac.uk/Tools/hmmer/ |
| RNAcentral nhmmer search | (RNAcentral Consortium, 2015; Bateman et al., 2011) | http://rnacentral.org/sequence-search/ |

## Alignment and Secondary Structure Prediction Tools

We find it best to run a variety of alignment and secondary structure prediction tools simultaneously. Each has its own peculiarities, and our hope is that by looking for shared homology and secondary structure predictions we can mitigate some of the problems discussed in the introduction. In this protocol, we use the WAR webserver (Torarinsson and Lindgreen, 2008) which allows the user to run 14 different methods simultaneously. These include Sankoff-type methods: FoldalignM (Torarinsson et al., 2007), LocARNA (Will et al., 2007), MXSCARNA (Tabei et al., 2008), Murlet (Kiryu et al., 2007), and StrAL (Dalli et al., 2006) with PETcofold (Seemann et al., 2011); Align-then-fold methods, which use a traditional alignment tool (ClustalW (Larkin et al., 2007) or MAFFT (Katoh and Standley, 2013)) followed by structure prediction (RNAalifold (Bernhart et al., 2008) or Pfold (Knudsen and Hein, 2003)); Fold-then-align methods, which predict structures in all the input sequences and attempt to align these structures (RNAcast (Reeder and Giegerich, 2005) + RNAforester (Höchsmann et al., 2003)); Sampling methods which attempt to iteratively refine alignment and structure: MASTR (Lindgreen et al., 2007) and RNASampler (Xu et al., 2007); and other methods which do not fit into the above traditional categories: CMfinder (Yao et al., 2006) and LaRA (Bauer et al., 2007). Finally, WAR also computes a maximum consistency alignment using all the alignment predictions with T-Coffee (Notredame et al., 2000) (Notredame et al., 2000).

However, WAR is by no means exhaustive, and the methods may not be the most recent versions available. A number of groups maintain their own servers for RNA sequence analysis. Notable servers include the Vienna RNA WebServers (Gruber et al., 2008b), the Freiburg RNA Tools (Smith et al., 2010), the CBRC Functional RNA Project (Asai et al., 2008; Mituyama et al., 2009), and the Center for Non-Coding RNA in Technology and Health (RTH) Resources page. In addition, EMBL-EBI maintains a number of webservers for popular multiple sequence alignment alignment tools (McWilliam et al., 2013). Ultimately, as you become more comfortable with RNA sequence analysis you may want to begin installing and running new tools on a local *NIX machine; however, this is beyond the scope of the current protocol.

| **Resource** | **Reference** | **URL** |
|---|---|---|
| Webserver for Aligning structural RNAs (WAR) | (Torarinsson and Lindgreen, 2008) | http://genome.ku.dk/resources/war/ |
| Vienna RNA | (Gruber et al., 2008b) | http://rna.tbi.univie.ac.at/ |
| Freiburg RNA Tools | (Smith et al., 2010) | http://rna.informatik.uni-freiburg.de |
| CBRC Functional RNA Project | (Asai et al., 2008) | http://software.ncRNA.org |
| RTH Resources | N/A | http://rth.dk/pages/resources.php |
| EMBL-EBI Alignment Tools | (McWilliam et al., 2013) | http://www.ebi.ac.uk/Tools/msa/ |

**Genome Browsers**

Genome browsers are essential for checking the context of putative homologs. The ENA (Silvester et al., 2015) provides a no-frills sequence browser perfect for quickly checking annotations. For deeper annotations, the UCSC genome browser (Karolchik et al., 2014) and Ensembl (Flicek et al., 2014) both contain a wide range of information for the organisms they cover. For bacterial and archaeal genomes, the Lowe lab maintains a modified version of the UCSC genome browser (Chan et al., 2012; Schneider et al., 2006) which provides a number of tracks of particular interest to those working with ncRNA. The CBRC Functional RNA Project maintains a UCSC genome browser mirror (Mituyama et al., 2009) for a number of eukaryotic organisms with a larger number of ncRNA-related tracks. Finally, the ENCODE Project (ENCODE Project Consortium, 2011) provides access to a large number of functional assays in a number of eukaryotic organisms, many of which have relevance to ncRNA. Offline genome browsers are also available for use on your own computer, for example IGV and Artemis (Thorvaldsdóttir et al., 2013; Carver et al., 2012). There are also methods available for comparing synteny (gene order) information between genomes (Alikhan et al., 2011; Sullivan et al., 2011; Carver et al., 2005).

| **Resource** | **Reference** | **URL** |
|---|---|---|
| European Nucleotide Archive | (Silvester et al., 2015) | http://www.ebi.ac.uk/ena |
| UCSC Genome Browser | (Karolchik et al., 2014) | http://genome.ucsc.edu/ |
| Ensembl | (Flicek et al., 2014) | http://www.ensembl.org |
| UCSC Microbial Genome Browser | (Chan et al., 2012; Schneider et al., 2006) | http://microbes.ucsc.edu/ |
| CBRC UCSC Genome Browser for Functional RNA | (Mituyama et al., 2009) | http://www.ncrna.org/glocal/cgi-bin/hgGateway |
| ENCODE Project | (ENCODE Project Consortium, 2011) | https://www.encodeproject.org/ |

**Alignment Editors**

It is possible to edit alignments in any text editor; however we highly recommend using a secondary structure-aware editor such as Emacs with the RALEE major mode (Griffiths-Jones, 2005). RALEE allows you to color bases according to base identity, secondary structure, or base conservation. It also allows the easy manipulation of sequences which are involved in structural interactions but are not close in sequence space through the use of split screens. A number of other specialized RNA editors are available: BoulderALE (Stombaugh et al., 2011) and S2S (Jossinet and Westhof, 2005) both allow the end user to visualize and manipulate tertiary structure in addition to secondary structure, and may be particularly useful if crystallographic information is available for your RNA. Other alternatives for editing RNA secondary structure are SARSE (Andersen et al., 2007) and MultiSeq (Roberts et al., 2006). Recent versions of JalView (Waterhouse et al., 2009) have begun to support RNA secondary structure as well. Finally, a recent attempt has been made to "gamify" RNA alignment editing. This approach, called Ribo, samples poorly resolved regions of alignments, and feeds these regions to gamers who manually refine the alignments. These re-aligned regions can then be reinserted into the original alignment (Waldispühl et al., 2015).

| **Resource** | **Reference** | **URL** |
|---|---|---|
| BoulderALE | (Stombaugh et al., 2011) | http://boulderale.sourceforge.net/ |
| JalView | (Waterhouse et al., 2009) | http://www.jalview.org |
| MultiSeq | (Roberts et al., 2006) | http://www.scs.illinois.edu/schulten/multiseq/ |
| RALEE | (Griffiths-Jones, 2005) | http://sgjlab.org/ralee/ |
| Ribo | (Waldispühl et al., 2015) | http://ribo.cs.mcgill.ca/ |
| S2S | (Jossinet and Westhof, 2005) | http://bioinformatics.org/S2S/ |
| SARSE | (Andersen et al., 2007) | http://sarse.ku.dk/ |

**Infernal**

The centerpiece of our protocol is the Infernal package for constructing covariance models (CMs) from RNA multiple alignments (Nawrocki and Eddy, 2013). We will use this to construct models of our RNA family. CMs model the conservation of positions in an alignment similar to a hidden Markov model (HMM), while also capturing *covariation* in structured regions (Eddy and Durbin, 1994; Sakakibara et al., 1994; Durbin et al., 1998). Covariation is the process whereby a mutation of a single base in a hairpin structure will lead to selection in subsequent generations for compensatory mutations of its structural partner in order to preserve canonical base-pairing, ie: Watson-Crick plus G-U pairs, and a functional structure. This combination of structural-evolutionary information has been shown to provide the most sensitive and specific homology search for RNA of any tools currently available (Freyhult et al., 2007; Gardner, 2009). Unfortunately, this sensitivity and specificity come at a high computational cost, and Infernal searches can be time-consuming with genome-scale searches often taking hours on desktop computers. The development of heuristics to reduce this computational cost is an area of active research for the Infernal team, and has already been mitigated to some extent by the use of HMM filters and query-dependent banding of alignment matrices (Nawrocki and Eddy, 2007, 2013). We refer the reader to Eric Nawrocki's primer on annotating functional RNAs in genomic sequence for a friendly introduction to the mechanics of the Infernal package (Nawrocki, 2014).

| **Resource** | **Reference** | **URL** |
|---|---|---|
| Infernal | (Nawrocki et al., 2009; Nawrocki and Eddy, 2013) | http://eddylab.org/software/infernal/ |

## Basic Protocols

We assume for the sake of this protocol that you are starting with a single sequence of interest. We will illustrate our protocol using the example of MicA, an Hfq-dependent bacterial trans-acting antisense small RNA (sRNA). Many bacterial sRNAs are similar in function to eukaryotic microRNAs, pairing to target mRNA transcripts through a short antisense-binding region, generally targeting the transcript for degradation (Barquist and Vogel, 2015). MicA is known to target a wide-range of outer membrane protein mRNAs using a 5′ binding-region in both *E. coli* (Gogol et al., 2011) and *S. enterica* (Vogel, 2009) in response to membrane stress. The previous covariance model for MicA (accession RF00078) in Rfam (release 10.1) was largely restricted to *E. coli*, *S. enterica*, and *Y. pestis*. Here, as an example, we produce an improved model, which has served as the core for the current Rfam model (release 12.0). In the process, we discover previously unreported MicA homologs in the nematode symbionts of the γ-proteobacterial genus *Xenorhabdus*.

If you already have a set of putative homologs, you may wish to further diversify your collection of sequences using the methods described in protocol 1, or you may skip directly to protocol 2, or 3 if a secondary structure is known. No matter how many sequences you are starting with, it is always a good idea to run the sequence search tools available on the Rfam website (http://rfam.xfam.org/) on them. This will verify that there isn't already a CM available that covers your sequences. There are a number of other specialist databases that may also be worth searching if you have reason to believe your RNA sequence is a member of a well-defined class of RNAs, i.e. microRNAs, snoRNAs, rRNAs, tRNAs, etc. We have recently reviewed these specialist RNA databases in (Hoeppner et al., 2014). A centralized RNA sequence database aiming to capture all known RNA sequences, RNAcentral (Bateman et al., 2011; RNAcentral Consortium, 2015), has recently launched and provides an integrated resource for easily identifying similar sequences with some evidence of transcription via an integrated nhmmer search.

### Protocol 1: Gathering an initial set of homologous sequence

Now that you have confirmed that your sequence is novel, we will use NCBI-BLAST to identify additional homologous sequences. Once you have navigated to the nucleotide BLAST server (http://blast.ncbi.nlm.nih.gov/) there are a number of important options to set.

**1.Setting NCBI-BLAST Parameters**

First, it is important to choose a search set appropriate to your sequence. At this initial phase, we want to limit our exposure to sequences which are very distant from ours to limit the number of obviously spurious alignments we will need to examine, increasing the power of our search. So, if your initial sequence is of human origin, you may want to limit your search to Mammalia, Tetrapoda, or Vertebrata depending on sequence conservation. Similarly, if you are working with an *Escherichia coli* sequence, you may want to limit your initial searches to Enterobacteriaceae or the Gammaproteobacteria. NCBI-BLAST searches are relatively fast, so try several search sets to get a feel for how conserved your sequence is.

The second set of options to set are the “Program Selection” and the “Algorithm Parameters”. We recommend **blastn** as it allows for smaller word sizes. The word size describes the minimum length of an initial perfect match needed to trigger an alignment between our query sequence and a target. Smaller word sizes provide greater sensitivity, and seem to perform better for non-coding RNAs. We recommend a word size of 7, the smallest the NCBI-BLAST server allows.

Finally, you should set “Max Target Sequences” parameter to at least 1000. NCBI-BLAST returns hits in a ranked list from best match to worst by E-value (or the number of matches with the same quality expected to be found in a search over a database of this size), and will only display as many as “Max Target Sequences” is set to. We are primarily interested in matches on the edge of what NCBI-BLAST is capable of detecting reliably, and these will naturally fall towards the end of this list.

Our example sequence, MicA, is from *E. coli*, so we will limit our initial searches to Enterobacteriaceae. This sequence, obtained from Gisela Storz’s collection of known *E. coli* sRNAs (http://cbmp.nichd.nih.gov/segr/ecoli_rnas.html) is:

```
GAAAGACGCGCATTTGTTATCATCATCCCTGAATTCAGAGATGAAATTTTGGCCACTCACGAGTGGCCTTTTT
```

Paste this sequence into the query sequence box at the top of the page. Limit the space of your search by entering “Enterobacteriaceae” (or equivalently “taxid:91347”) in the “Organism” box. Under “Program Selection”, click the radio button to select **blastn** search. Access additional options by clicking the plus sign next to “Algorithm parameters”. Under “General Parameters” set “Max target sequences” to 1000 and “Word size” to 7. Click the “BLAST” button to run the search.

## 2. Selecting Sequences

Our goal at this stage is to pick a representative set of homologous sequences to “seed” our alignment with. As discussed in the introduction, single sequence alignment for nucleotides is generally only reliable to around 60 percent pairwise sequence identity. At the same time, picking a large number of sequences with high percent identity can lead to *overfitting* of the secondary structure; that is, if our sequences are too similar we can end up predicting alignments and secondary structures which capture accidental features of a narrow clade, rather than conserved structure and sequence variation.

**FIGURE 1 APPROX HERE**

There are 3 major criteria we pick additional sequences based on, in rough order of importance: percent sequence identity, taxonomy, and sequence coverage. Handily, the NCBI-BLAST output displays measures of all of these (Figure 1). Our first selection criterion, percent identity, should fall between 65% and 95%: much lower and the sequence will be difficult to align, higher and it will be too similar to have any meaningful variation.

The second selection criterion, taxonomy, will depend somewhat on the organisms your sequence is associated with, but we generally want to limit the inclusion to a single (orthologous) instance per species. The exception to this rule is for diverged paralogous sequences within the species; if paralogs exist, you will need to decide how broadly you wish to define your family, though it can be difficult to construct families that capture the full phylogenetic range of an element while retaining the ability to discriminate between paralogs within single genomes. Additionally, it may be useful to further limit the maximum percent identity to, say, 90% within a densely sequenced genus to further limit the number of highly similar sequences in your initial alignment.

Finally, assuming that you are sure of your sequence boundaries, we want to select sequences that cover the entire starting sequence. If you see many matches covering only a short section of your sequence, this may be due to the matching of a short convergent motif. This most commonly happens with the relatively long, highly-constrained bacterial rho-independent terminators (Gardner et al., 2011), but may occur with other motifs (Gardner and Eldai, 2015). Alternatively, if you do not have well-defined sequence boundaries, you will need to determine these from the conservation you see in your BLAST hits - look for taxonomically diverse hits covering the same segment of your query sequence. In some cases, such as the long non-coding RNAs, conserved domains may be much shorter than the complete transcribed sequence (Burge et al., 2013), but stay aware of the potential motif issue. A taxonomic distribution of sequences that makes biological sense given your knowledge of the molecule's function and that can be explained by direct inheritance of the sequence will be your best guide.

Continuing our MicA example, we want to select a group of sequences with a reasonably diverse taxonomic range and as much sequence diversity as possible, while being reasonably confident that they are true homologs. In this case we will choose based on maximizing genus diversity, a percent identity between 75% and 90%, and 100% sequence coverage as we're fairly confident in the MicA gene boundaries. You can retrieve individual sequences by clicking “Download” on the alignment, then selecting the “FASTA (aligned sequence)” option. Note that you may have to reverse complement sequences (for instance using this webserver: http://www.bioinformatics.org/sms/rev_comp.html), as by default NCBI-BLAST returns sequences from the forward strand of the genomic sequence. For our example alignment, in addition to our original sequence from E. coli (EMBL-Bank accession: U00096) we have chosen sequences from *Salmonella* Typhimurium (FQ312003), *Klebsiella pneumoniae* (CP002910), *Enterobacter cloaca* (CP002272), *Yersinia pestis* (AM286415), *Pantoea* sp. At-9b (CP002433), and *Erwinia pyrifoliae* (FP236842). Collect these in FASTA format as below:

```
>U00096.3
GAAAGACGCGCATTTGTTATCATCATCCCTGAATTCAGAGATGAAATTTTGGCCACTCACGAGTGGCCTTTTT
>FQ312003
GAAAGACGCGCATTTGTTATCATCATCCCTGTTTTCAGCGATGAAATTTTGGCCACTCCGTGAGTGGCCTTTTT
>CP002272
GAAAGACGCGCATTTGTTATCATCATCCCTGACTTCAGAGATGAAATGTTTGGCCACAGTGATGTGGCCTTTTT
>CP002910
GAAAGACGCGCATTTATTATCATCATCATCCCTGAATCAGAGATGAAAGTTTGGCCACAGTGATGTGGCCTTTTT
>AM286415
GAAAGACGCGCATTTGTTATCATCATCCCTGTTATCAGAGATGTTAATTTGGCCACAGCAATGTGGCCTTTT
>CP002433
GAAAGACGCGCATTTGTTATCATCATCCCTGACAACAGAGATGTTAATTCGGCCACAGTGATGTGGCCTTTT
>FP236842
GAAAGACGCGTATTTGTTATCATCATCTCATCCCTGACAACAGAGATGTTAATTTAGGCCACAGTGACGTGGCCTTTTT
```

### 3. Examining Your Initial Homolog Set

Once you have assembled a set of sequences fitting the criteria described above, it is worth taking a closer look at them. Remember that these sequences will form the core of your alignment and CM, and errors at this stage can dramatically bias your results. A good first test is to examine the taxonomy of your sequences, and make sure it makes sense. Can you identify a clear pattern of inheritance that might explain the taxonomic distribution you see at this stage? Another good check is to examine your sequences in the ENA browser, or a domain-specific browser if one exists for your organisms. For many independently transcribed RNAs, genomic context is better conserved than sequence, and ncRNA genes will often fall in homologous intergenic or intronic regions even over large evolutionary distances. If you are particularly ambitious, and the tools are available for your organisms of interest, you may wish to try to identify promoter sequences upstream of your candidate or terminator sequences downstream. If your sequence is a putative cis-regulatory element, such as a riboswitch, thermosensor, or attenuator, you may want to check that it occurs upstream of genes with similar functions or in similar pathways (Weinberg et al., 2015). Finally, it is always worth searching your putative homologs through the Rfam website even if your initial sequence had no matches - Rfam's models are not perfect, and may miss distant homologs of known families.

**FIGURE 2 APPROXIMATELY HERE**

You can quickly examine your chosen sequences with the ENA browser, accessed by appending the accession number to http://www.ebi.ac.uk/ena/data/view/ (Figure 2). It appears that all of these sequences fall in a intergenic region between a *luxS* protein homolog and a *gshA* protein homolog, further increasing our confidence that these are true homologs. From our results, we can also see a few promising hits that don't quite meet our criteria, such as *Dickeya*, *Xenorhabdus*, *Photorhabdus*, and *Wigglesworthia*. We will keep these in mind later to expand our coverage.

## Protocol 2: Aligning and predicting secondary structure

We will use the WAR server to construct an initial alignment. Because of the criteria we've set for sequence similarity in our gathering step, all of the sequences in our initial homolog set should have at least 60% pairwise sequence identity with at least one other sequence in the set. Under these conditions alignment methods using primary sequence information only can perform adequately, as discussed in the background section. These methods combined with alignment folding tools that identify conserved structural signals and covariation can produce reasonable secondary structures predictions (Gardner and Giegerich, 2004). However it is still often useful to observe the behavior of as many alignment tools as possible. Using WAR, for a fairly fast alignment we recommend running CMfinder, StrAL+PETfold, ClustalW and MAFFT with RNAalifold and Pfold. WAR will also produce a consensus alignment using T-Coffee, which will attempt to find an alignment consistent with all of the individual alignments produced by other methods.

### 1. Submit your sequences to the WAR webserver

You will need to format your input sequence in FASTA format, an example can be found in protocol 1.2. Navigate to http://genome.ku.dk/resources/war/. Paste this into the sequence box. Uncheck the tick boxes next to all methods besides CMfinder, Stral+PETfold, ClustalW+Pfold, ClustalW+RNAalifold, MAFFT+Pfold, MAFFT+RNAalifold; this will limit WAR to running relatively fast methods. You can include additional methods in later runs to see how this affects the resulting alignments. Submit this together with your email address. Using fast alignment methods you should get results within 10 minutes, assuming the server isn't busy.

**FIGURE 3 APPROXIMATELY HERE**

**2. Understanding WAR results**

Once WAR returns your alignment results, there are a number of things you should take note of that will assist you in picking an alignment and further in manual refinement. First, the consensus alignment page will display a graphical representation of the consistency of the alignments which will allow you to quickly tell which areas of the alignment may require attention during manual refinement, or areas that may harbor structure not captured by the majority consensus. For instance, in our example MicA alignment (Figure 3) there is some alignment and structural disagreement around the first alignment gap, suggesting this area might require some manual curation to correct. The consensus can be recomputed based on differing subsets alignment methods, if you believe one method (or set of methods) may be unduly influencing the consensus. Once you've carefully looked over the consensus alignment, examine each alignment produced by WAR in turn: What structures are shared? Where do the alignments differ from each other? Can you identify  any sequence or structural motifs which may help to guide your alignment? At this level of sequence identity, you should hope to see fairly consistent alignments in functional regions of the sequence, interspersed with more difficult to align regions, presumably under weaker selective pressure. Often the consensus alignment is a good choice to move forward with. However, there are cases where certain classes of tools will obviously mis-align regions of the sequence and bias the consensus. Keep in mind what you've seen in the alternative alignments as well; this information may be useful in manual refinement. You will want to save the stockholm file for the alignment you've chosen to your local computer at this point. This can be done by clicking the "Stockholm" link at the top right of any result page.

**3. Aligning more distantly related sequences**

Later in the family-building process when you have identified more distant homologs, the average pairwise identity of the sequences in your data set may have dropped below 60%. At this point, you may want to begin including some of the Sankoff-type alignment methods available in WAR. Using these methods can dramatically increase the runtime for your sequence alignment jobs, though, particularly for sequences over a couple of hundred of bases long. We will discuss alternatives to re-aligning sequences during the iterative expansion of the alignment in section 3.5.

## Protocol 3: Guidance for manually refining alignments

Our goal in manual refinement is to attempt to correct errors made by automatic alignment tools. We generally use RALEE (Griffiths-Jones, 2005), an RNA editing mode for Emacs, for editing alignments. However, any editor you are comfortable with in which you can easily visualize sequence and structural conservation will work; a number of alternative editors are listed in the strategic planning section. RALEE offers coloring based on stucture, sequence conservation, and compensatory mutations. These can be accessed from the “Structure” menu in Emacs (Figure 4).

**FIGURE 4 APPROXIMATELY HERE**

A good place to start editing is around the edges of predicted hairpin structures. Are there base-pairs which appear to be misaligned? Can you add base-pairs to the structure? Are there predicted base-pairs which don't appear to be well conserved that should be trimmed? Can individual bases be moved in the alignment to create more convincing support for the predicted structure?

Once you are satisfied with your manual refinement of predicted secondary structure elements, next you should turn your attention to areas identified as uncertain in the WAR/T-Coffee consensus alignment. Were there alternative structures predicted in these regions? Do you see support for these structures in the sequences? If these regions are unstructured, can you identify any conserved sequence motifs in the region? If you will be regularly working with a particular class of ncRNA, it can be useful to familiarize yourself with predicted binding motifs of associated RNA-binding proteins as these are likely to be conserved but can have many variable positions (Gardner and Eldai, 2015).

Our MicA consensus alignment provides a few opportunities for improvement (Figure 5). The first base-pair in the first stem in CP002433 can be rescued by shifting a few nucleotides, and by pulling apart the alignment between the first and second stem we reveal what appears to be a well-conserved AAUUU sequence motif that was previously hidden (Figure 5). The bacterial RNA chaperone Hfq is known to bind to A/U rich sequences, so this motif may have some functional significance (Schu et al., 2015).

**FIGURE 5 APPROXIMATELY HERE**

At this stage, it is also possible to include information from experimental data. Crystal structure information from a single sequence in the seed alignment can be used to validate and improve a predicted secondary structure. Tertiary structure-aware editors such as BoulderAle (Stombaugh et al., 2011) can help in applying this information to the alignment. Other experimental evidence, such as chemical footprinting, particularly when coupled with high-throughput sequencing (Spitale et al., 2014; Kwok et al., 2015), can also provide valuable information. Knowing whether even a single base is involved in a pairing interaction can drastically reduce the space of possible structures the sequence can fold into, simplifying the problem of predicting secondary structure. Both the RNAfold and RNAalifold web servers available through the Vienna RNA website (Gruber et al., 2008b) are capable of taking advantage of this information in the form of folding constraints. Structural mapping datasets are beginning to be archived in consistent formats (Rocca-Serra et al., 2011; Cordero et al., 2012), and will be increasingly available for a variety of non-coding RNA molecules in the future.

## Protocol 4: Building a covariance model

For those comfortable with the *NIX command line, building an Infernal CM is fairly straight-forward. We refer the reader to the User's Guide available from the Infernal website (http://http://eddylab.org/software/infernal/) for installation instructions and a detailed tutorial.

### 1. Run cmbuild

Assuming you have successfully installed Infernal somewhere in your path and you have saved your edited alignment in a stockholm file named my.sto (see figure 5 for an example of stockholm format), run:

```
> cmbuild my.cm my.sto
```

This will construct a CM and save the results in a file named my.cm. This command also returns some useful statistics on your model, such as the effective number of sequences, and the relative entropy of your CM compared to an HMM with the same sequences, i.e. how much information is gained by including structural information in your model.

### 2. Run cmcalibrate

To generate E-values in search results, the CM must now be calibrated. This calibration depends on a sampling procedure, and so can be time-consuming. For our MicA CM, it will take about 5 minutes. Run:

```
> cmcalibrate my.cm
```

Note that calibration can take a long time -- hours for longer models. You can get a quick estimate of the time calibration will take using the command:

```
> cmcalibrate --forecast 1 my.cm
```

Congratulations! You should now have a working CM for your RNA family. This is a fully capable model, and can be used as is for homology search and genome annotation. However, as it stands, your CM will only capture the sequence diversity which was able to be detected by our initial BLAST search. In order to fully take advantage of the power of CMs, you may want to expand the diversity of the sequence it is trained on through iterative expansion of our initial set of sequence homologs.

## Protocol 5: Strategies for expanding model coverage

### Plan A: Iterative search of sequence databases

The method Rfam used to identify more divergent homologs to seed sequences prior to recent HMM-based acceleration of the Infernal pipeline (Nawrocki and Eddy, 2013; Nawrocki et al., 2015) was to pre-filter CM-based searches with WU-BLAST. This allows us to cover a large sequence space with a (comparatively) modest investment of computational time. Any of the single sequence search tools mentioned in the strategic planning section would make an effective pre-filter.

The easiest way to perform filtering yourself is to use the NCBI BLAST webserver to search each sequence in your seed alignment following the methods outlined for collecting your initial set of homologs in protocol 1. You may wish to relax the criteria slightly, then use the CM to perform a more sensitive search on this set of filtered sequences. This will enable you to detect more distantly related sequences, though you should always examine sequence context and the phylogenetic relationship between sequences as a sanity check before including them in your seed. These methods can be automated with basic scripting and bioinformatics modules such as BioPerl (Stajich et al., 2002) or Biopython (Cock et al., 2009), though this is beyond the scope of this chapter.

Once you have identified a new set of homologs, you can align them to your previous CM using Inferal's cmalign:

```
> cmalign --mapali my.sto my.cm newsequences.fasta >
mynewalignment.sto
```

This `--mapali` option includes your original training alignment in the `mynewalignment.sto` output. This alignment can then be used to build a new CM, which will capture the additional sequence variation you have discovered in your BLAST searches.

The disadvantage of this method is that each search only uses the information available in a single sequence, meaning that valuable information about variation is lost and as a result the power of the search suffers. Fast profile-based methods implemented in HMMER3 (Wheeler and Eddy, 2013) can be used to remedy this (Lindgreen et al., 2014), but require significant local computational resources to be used in a manner similar to the NCBI-BLAST.

**Plan B: Directed search of chosen sequences**

Another approach is to run the unfiltered CM over selected genomes or genomic regions. While the greater sensitivity and specificity of this method can help identify more distant homologs than is possible with BLAST, it has the disadvantage that it requires a much larger investment of computational resources to provide an equivalent phylogenetic coverage. This method can be particularly powerful in bacterial and archaeal genomes, where small genome size allows us to search a phylogenetically-representative sample of genomes in less than a day on a desktop computer. In the case of larger eukaryotic genomes, it may be necessary to search a few genomes to determine if homologs of your RNA are likely to exist in certain lineages, then extract homologous intergenic regions to continue searching. Our rationale here is much the same as in limiting the database for our initial BLAST search: by only looking in genomes where we have some prior belief that they may contain homologous sequence we reduce the noise in our low-scoring hits, meaning that we have to manually examine less hits to establish a score threshold for likely homologs.

Once you have examined candidates following the principles outlined earlier, it is easy to incorporate your new sequences using the easel package included with Infernal. First, search the genome generating a tabfile:

```
> cmsearch --tblout searchfile.tab my.cm genome.fasta
```

Then use easel to index the genome and extract the hits:

```
> esl-sfetch --index genome.fasta
> esl-sfetch -Cf searchfile.tab genome.fasta > hits.fasta
```

These sequences can then be aligned and merged as with BLAST hits. Alternatively, if you discover a divergent lineage, it may be easiest to construct a separate alignment for these sequences, then use shared structural and sequence motifs to manually combine the two alignments. Sankoff-type alignment method may also be useful for aligning divergent clades.

**Plan C: When A and B fail...**

In some cases, it will be very difficult to identify homologs of a candidate RNA across its full phylogenetic range. This can be because of high sequence variability, as in the Vault RNAs (Stadler et al., 2009). Alternatively, some longer RNAs, such as the RNA component of the telomerase ribonucleoprotein, consist of structurally conserved segments interspersed with long variable regions that can't be easily discovered by standard search with naive covariance models (Marz et al., 2012; Xie et al., 2008).

A number of computational techniques exist for approaching these difficult cases, reviewed by Mosig and colleagues (Mosig et al., 2009). These methods include fragrep2 (Mosig et al., 2007), which allows the user to search fragmented conserved regions (including pol III promotor and terminator motifs (Stadler et al., 2009)(Gruber et al., 2008a)(Marz et al., 2009)), fragrep3, which allows the user to incorporate custom structural motifs with fragmented search, and GotohScan (Hertel et al., 2009), which implements a *semi-global* alignment algorithm that will align a query sequence to a (potentially) extended genomic region.

**Applying Plans A & B to MicA**

Now we will follow Plan B to add sequences to our alignment using the genomes for the low-scoring BLAST hits we had previously made a note of while collecting our initial set of sequences, though you could also choose these sequences based on your knowledge of your organisms phylogeny or the suspected function of your RNA. The genomes we have chosen here are *Dickeya zeae* (CP001655), *Sodalis glossinidius* (AP008232), *Xenorhabdus nematophila* (FN667742) and *Wigglesworthia glosinidia* (BA000021); these can all be downloaded from the EMBL bacterial genomes pages (http://www.ebi.ac.uk/genomes/bacteria.html). Searching these genomes allows us to identify strong hits in *D. zeae* and *S. glossinidius* with E-values of $10^{-12}$ and $10^{-10}$ which we can merge into our alignment using the methods described in Plan B above. You should then manually refine the resulting merged alignment with an eye towards maintaining conserved sequence motifs and structure. Already at this evolutionary distance, there has been some apparent small decay in secondary structure, as well as an expansion of the sequence contained in the loop region of the second stem in *D. zeae* (Figure 6).

**FIGURE 6 APPROXIMATELY HERE**

We observe a number of hits in *X. nematophila* with E-values in the range of $10^{-2}$, and we can apply Plan A to investigate these. By checking each of these individually in the ENA browser, we can identify one that falls in the same genomic context as our previous MicA homologs (Figure 7). By using this sequence as the starting point for a BLAST search, we can identify a number of other divergent *Xenorhabdus* homologs. As these are quite diverged from the *E. coli* sequence, we first construct an alignment for them using WAR (Figure 7), then attempt to merge our alignments manually using shared structural features as our guide. Interestingly, the target-binding region of MicA appears to have suffered a poly-A insertion along this lineage, suggesting that there may be changes in the regulon it targets, in line with recent findings that RNA:RNA interactions are frequently poorly conserved (Lai and Meyer, 2015; Richter and Backofen, 2012). Using this model to search all of the bacterial genomes in EMBL-Bank (approximately 6GB of sequence, taking ~30 hours on a 2.26 GHz Intel Core 2 Duo processor) shows that our CM now has high-scoring hits exclusively in Enterobacteriales, while covering a broader range than our initial BLAST searches. This search also reveals a number of possible sources of additional diversity: *Photorhabdus asymbiotica* and *Edwardsiella ictaluri* both have strong hits below the average score for other enterobacterial genomes -- incorporating them may further increase the sensitivity of our model, and is left as an exercise to the reader.

## Commentary

### Background Information

RNA sequence alignment remains a challenge despite at least 30 years of work on the problem (Woese et al., 1983; Lane et al., 1985; Guttel et al., 1985; Pace et al., 1989). As discussed in the introduction, alignments based on primary sequence become untrustworthy below ~60% pairwise sequence identity, likely due to the lower information content of individual nucleic acids as compared to amino acids in protein alignments. This can be intuitively understood by recalling the fact that 3 nucleic acids are required to encode an individual amino acid; so, an amino acid carries 3 times as much information as a nucleic acid (a bit less, actually, due to the redundancy of the genetic code). In addition, the larger alphabet size of protein sequences allows for the easy deployment of more complex substitution models, and a glut of protein sequence data allows for highly effective parameterization of these models.

The incorporation of secondary structure, i.e. base-pairing, information has been proposed as a means to make up for these difficulties in RNA alignment methods. The first proposal for such a method is now known as the Sankoff algorithm (Sankoff, 1985). The Sankoff algorithm uses dynamic programming, an optimization technique central to sequence analysis[1]. Dynamic programming had previously been applied to the problems of sequence alignment (Needleman and Wunsch, 1970) and RNA folding (Nussinov et al., 1978). Sankoff proposed a union of these two methods. Unfortunately, the resulting algorithm has a time requirements of $O(L^{3N})$ and space requirements of $O(L^{2N})$ where $L$ is the sequence length and $N$ is the number of sequences aligned. This is impractical, even for small numbers of short sequences. A number of faster heuristic algorithms have been developed to approximate Sankoff alignment. Recent examples include CentroidAlign (Hamada et al., 2009b), mLocARNA (Will et al., 2007), and FoldalignM (Torarinsson et al., 2007). These methods can push the RNA alignment twilight zone as low as 40 percent identity (Gardner et al., 2005).

[1] A full explanation of dynamic programming is beyond the scope of this chapter, but for a brief introduction see two excellent primers by Sean Eddy covering applications to alignment (Eddy, 2004b) and secondary structure prediction (Eddy, 2004a); for those seeking a deeper understanding (Durbin et al., 1998) provides detailed coverage of dynamic programming as well as covariance models.

However, for the purpose of family-building, we are often starting with a single sequence of unknown secondary structure, and have to gather additional homologs using a fast alignment tool, such as BLAST. This is a common starting point in analysing non-coding coding regions from RNA sequence experiments, for example (Perkins et al., 2009; Chaudhuri et al., 2011; Lindgreen et al., 2014). Unfortunately, BLAST and similar single-sequence homology search methods are not able to reliably detect homologs below 60 percent sequence identity. In this range of pairwise sequence identities, the slight increases in accuracy of Sankoff-type algorithms over non-structural alignment is only rarely worth the additional computational costs involved[2]. Alignments generated with standard alignment tools can then be used as a basis for predictions of secondary structure using tools like Pfold (Knudsen and Hein, 2003), RNAalifold (Bernhart et al., 2008), or CentroidFold (Hamada et al., 2009a).

Regardless, all modern alignment tools, Sankoff-type or standard, suffer from a number of known problems. Most alignment tools use *progressive alignment*. This means that the aligner decomposes the alignment problem into a series of pair-wise alignment problems along a guide tree built using some measure of sequence similarity, so that highly similar sequences are aligned first before alignments between more divergent sequences are attempted. This greatly reduces the computational complexity of the alignment problem, but also means that errors in early pairwise alignment steps are propagated through the entire alignment. A number of solutions have been proposed to this problem, such as explicitly modeling insertion-deletion histories (Löytynoja and Goldman, 2008) or using modified or alternative optimization methods such as consistency-guided progressive alignment (Notredame et al., 2000) or sequence annealing (Schwartz and Pachter, 2007). A second issue is that it is not clear which function of the alignment aligners should be optimizing (Iantorno et al., 2014), and many appear to over-predict homology (Schwartz et al., 2005; Bradley et al., 2008, 2009). Finally, many parameters commonly used in alignment, such as gap opening and closing probabilities and substitution matrices, appear to vary across organisms, sequences, and even positions within an alignment. All of this leads to considerable uncertainty in alignment (Wong et al., 2008), which is not easily captured by most current alignment methods, although sampling-based methods do exist which attempt to account for this (Suchard and Redelings, 2006; Westesson et al., 2012). The additional parameters and uncertainty introduced by RNA secondary structure prediction only compounds these problems.

An additional problem with alignment is the issue of determining whether two sequences are similar due to *homology* or *analogy*. Homology describes a similarity in features based on common descent; for instance, all bird wings are homologous wings. Analogy, on the other hand, describes a similarity in features based on common function without common descent; bat and bird wings perform the same function and appear superficially similar, however, their evolutionary histories are quite different. In sequence analysis, we often assume that aligned residues within an alignment share common ancestry, but this assumption can be confounded by analogous sequence. These analogs often take the form of *motifs*, short sequences which perform specific functions within the RNA molecule and can arise easily through convergent evolution (Gardner and Eldai, 2015). An example of such a motif is the bacterial rho-independent terminator (Gardner et al., 2011), a short hairpin responsible for halting transcription in many species. While such motifs can be a boon in discovering novel ncRNA genes (Argaman et al., 2001; Livny et al., 2005) or aligning homologs which contain them (Nawrocki et al., 2015), they can also be a source of false-positives when attempting to build an alignment of homologous sequences.

[2] For relatively recent benchmarks of alignment tools on ncRNA sequences see (Hamada et al., 2009b) and the supplementary information of (Bradley et al., 2008); Hamada includes comparisons of aligner runtimes, while Bradley examines relative performance over a range of pair-wise sequence percent identities.

| | | | | | | |
|---|---|---|---|---|---|---|
| Cronobacter sakazakii strain NCTC 8155, complete genome | 91.5 | 91.5 | 100% | 5e-17 | 89% | CP012253.1 |
| Cronobacter sakazakii strain ATCC 29544, complete genome | 91.5 | 91.5 | 100% | 5e-17 | 89% | CP011047.1 |
| Pluralibacter gergoviae strain FB2, complete genome | 91.5 | 91.5 | 100% | 5e-17 | 88% | CP009450.1 |
| Salmonella enterica subsp. enterica serovar Enteritidis str. EC20130347 genome | 91.5 | 91.5 | 79% | 5e-17 | 95% | CP007423.1 |
| Cronobacter sakazakii CMCC 45402, complete genome | 91.5 | 91.5 | 100% | 5e-17 | 89% | CP006731.1 |
| Cronobacter sakazakii SP291, complete genome | 91.5 | 91.5 | 100% | 5e-17 | 89% | CP004091.1 |
| Cronobacter sakazakii ES15, complete genome | 91.5 | 91.5 | 100% | 5e-17 | 89% | CP003312.1 |
| Cronobacter turicensis z3032 complete genome | 91.5 | 91.5 | 100% | 5e-17 | 89% | FN543093.2 |
| Cronobacter sakazakii ATCC BAA-894, complete genome | 91.5 | 91.5 | 100% | 5e-17 | 89% | CP000783.1 |
| Enterobacter sp. 638, complete genome | 91.5 | 91.5 | 100% | 5e-17 | 89% | CP000653.1 |
| Cronobacter muytjensii ATCC 51329, complete genome | 86.0 | 86.0 | 100% | 2e-15 | 88% | CP012268.1 |
| Kluyvera intermedia strain CAV1151, complete genome | 86.0 | 86.0 | 100% | 2e-15 | 86% | CP011602.1 |
| Klebsiella michiganensis strain RC10, complete genome | 82.4 | 82.4 | 76% | 3e-14 | 93% | CP011077.1 |
| Cedecea neteri strain ND14a, complete genome | 82.4 | 82.4 | 76% | 3e-14 | 93% | CP009459.1 |
| Cedecea neteri strain M006, complete genome | 82.4 | 82.4 | 76% | 3e-14 | 93% | CP009458.1 |
| Cedecea neteri strain SSMD04, complete genome | 82.4 | 82.4 | 76% | 3e-14 | 93% | CP009451.1 |
| Yersinia kristensenii strain Y231, complete genome | 80.6 | 80.6 | 98% | 1e-13 | 85% | CP009997.1 |
| Yersinia frederiksenii Y225, complete genome | 80.6 | 80.6 | 98% | 1e-13 | 85% | CP009364.1 |
| Pantoea rwandensis strain ND04, complete genome | 80.6 | 80.6 | 100% | 1e-13 | 87% | CP009454.1 |
| Yersinia kristensenii strain ATCC 33639 genome | 80.6 | 80.6 | 98% | 1e-13 | 85% | CP008955.1 |
| Shimwellia blattae DSM 4481 = NBRC 105725, complete genome | 80.6 | 80.6 | 100% | 1e-13 | 85% | CP001560.1 |
| Klebsiella pneumoniae strain J1, complete genome | 78.8 | 78.8 | 100% | 3e-13 | 84% | CP013711.1 |
| Klebsiella pneumoniae strain U25 genome | 78.8 | 78.8 | 100% | 3e-13 | 84% | CP012043.1 |

Figure 1: **Partial results of a BLAST search using the *E. coli* MicA sequence from the "nr" sequence database.** The tabular view provides important information that can be used to pick putative homolog sequences including species and strain information (column 1), query sequence coverage (column 4), E-value (column 5), and percent identity (column 6). Also note the accession number in column 7, this will be useful looking up sequences in other databases (e.g. ENA).

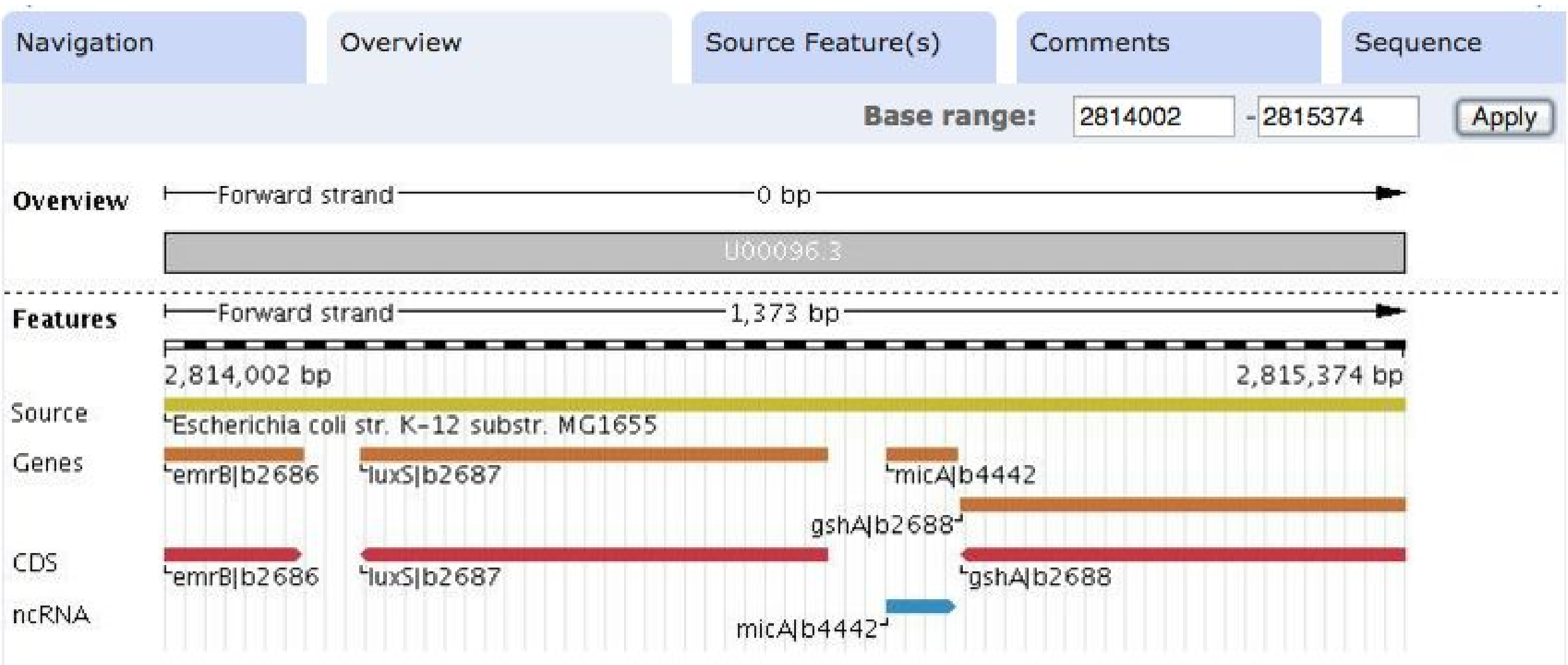

Figure 2: **Genomic context of *micA*.** ENA browser view of the region surrounding *micA* in *E. coli*. All of our selected homologs show conserved synteny with the *luxS* and *gshA* genes, providing additional evidence for their evolutionary relationship. Note the Rfam annotation in this region, matching our sequence. This view can be generated by navigating to http://www.ebi.ac.uk/ena/data/view/U00096.3 and entering the genome coordinates in the "Base range" boxes.

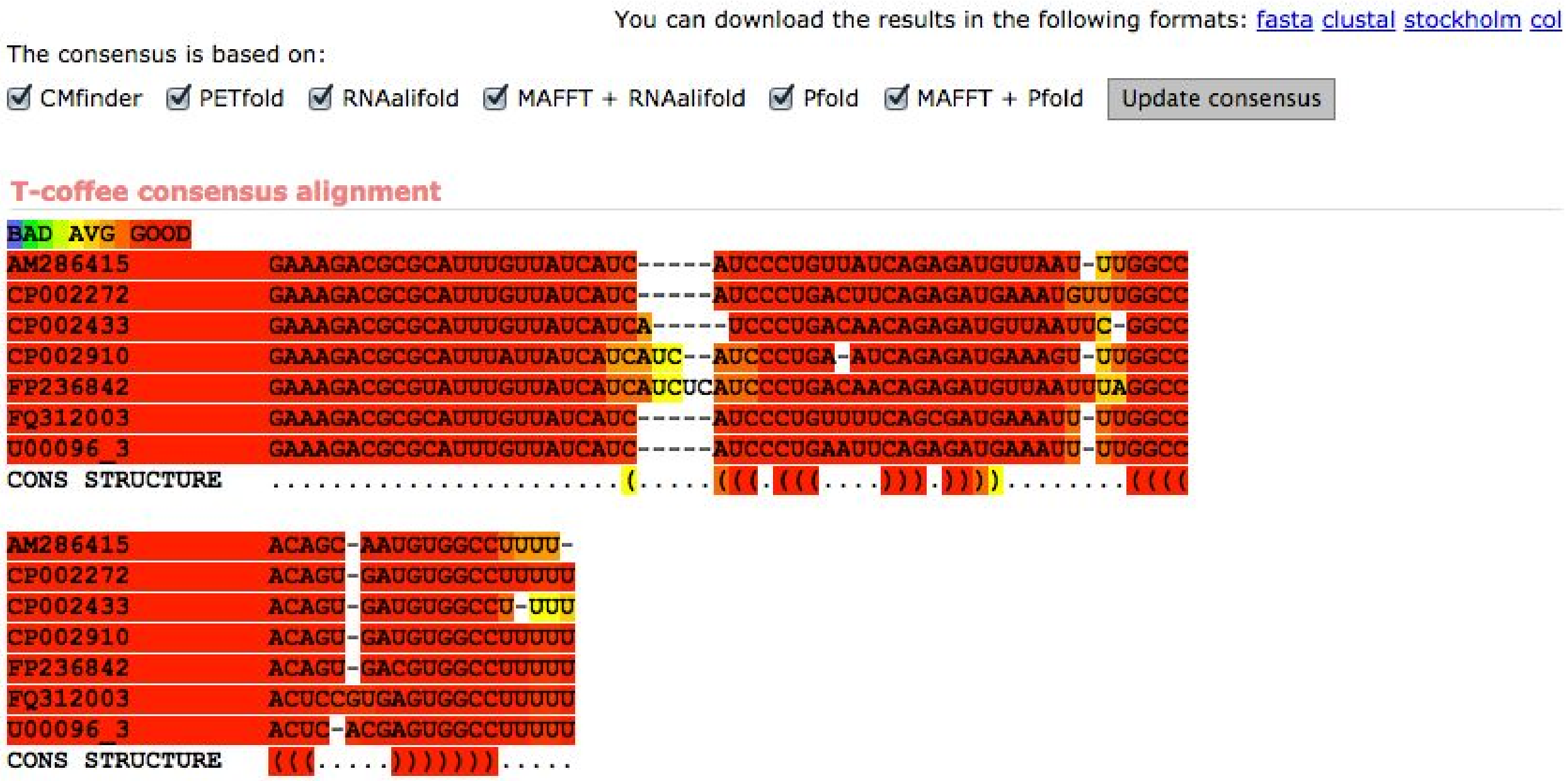

Figure 3: **Consensus alignment of MicA sequences.** Visualization of a T-coffee consensus alignment incorporating information from 6 different alignment and structure prediction methods on the WAR webserver.

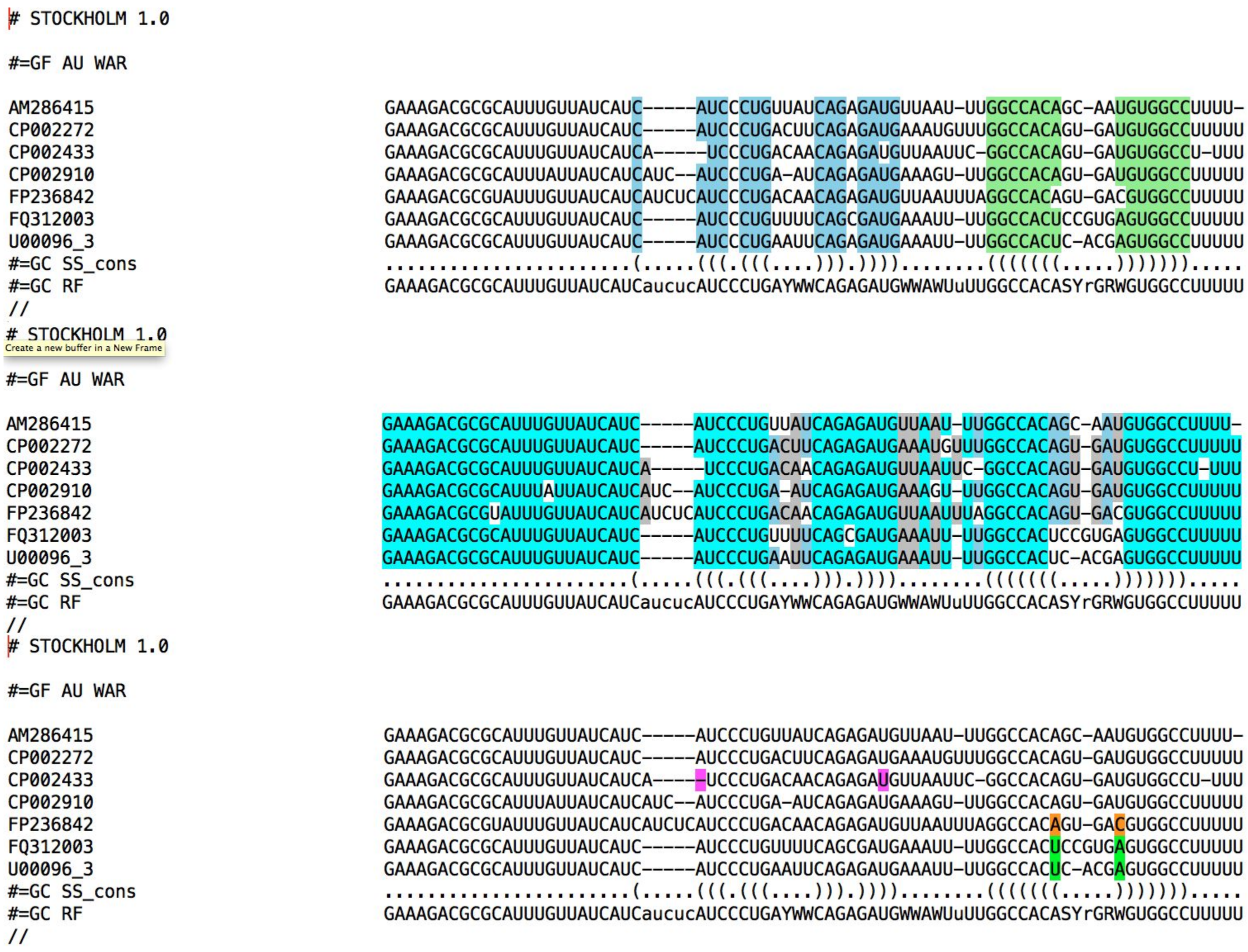

Figure 4: **Editing alignments in emacs RALEE mode.** Screenshots showing three different colour markups of the MicA consensus alignment highlighting secondary structure (top), sequence conservation (middle), and compensatory mutations (bottom).

```
#=GF AU WAR

AM286415                  GAAAGACGCGCAUUUGUUAUCAU.....CAUCCCUGUUAUCAGAGAUGUUAA...UUU.GGCCACAGC-AAUGUGGCCUUUU-
CP002272                  GAAAGACGCGCAUUUGUUAUCAU.....CAUCCCUGACUUCAGAGAUG..AAAUGUUU.GGCCACAGU-GAUGUGGCCUUUUU
CP002433                  GAAAGACGCGCAUUUGUUAUCAU.....CAUCCCUGACAACAGAGAUGUUAA....UUCGGCCACAGU-GAUGUGGCCU-UUU
CP002910                  GAAAGACGCGCAUUUAUUAUCAUCAU..CAUCCCUGA-AUCAGAGAUG..AAA.GUUU.GGCCACAGU-GAUGUGGCCUUUUU
FP236842                  GAAAGACGCGUAUUUGUUAUCAUCAUCUCAUCCCUGACAACAGAGAUGUUAA...UUUAGGCCACAGU-GACGUGGCCUUUUU
FQ312003                  GAAAGACGCGCAUUUGUUAUCAU.....CAUCCCUGUUUUCAGCGAUG..AAAU.UUU.GGCCACUCCGUGAGUGGCCUUUUU
U00096_3                  GAAAGACGCGCAUUUGUUAUCAU.....CAUCCCUGAAUUCAGAGAUG..AAAU.UUU.GGCCACUC-ACGAGUGGCCUUUUU
#=GC SS_cons              ............................((((.(((....))).))))..........(((((((.....)))))))).....
#=GC RF                   GAAAGACGCGCAUUUGUUAUCAUcaucuCAUCCCUGAYWWCAGAGAUGWW..AWUuUU.GGCCACASYrGRWGUGGCCUUUUU
//
```

Figure 5: **MicA alignment following manual refinement.** Manual refinement using RALEE restores full conservation to the first stem-loop and removes an unlikely insertion within the hairpin structure. Additionally careful manipulation of the sequence between the two hairpin reveals strong conservation of a short A/U-rich sequence motif that was previously obscure.

```
# STOCKHOLM 1.0

U00096.2        GAAAGACGCGCAUUUGUUAUCA.....UCAUCCCUGAAUUCAGAGAUG..AAAU.UUU.GGCCAC.U..CA.CG.....AGUGGCCUUUUU
FQ312003        GAAAGACGCGCAUUUGUUAUCA.....UCAUCCCUGUUUUCAGCGAUG..AAAU.UUU.GGCCAC.U..CCgUG.....AGUGGCCUUUUU
CP002272        GAAAGACGCGCAUUUGUUAUCA.....UCAUCCCUGACUUCAGAGAUG..AAAUgUUU.GGCCAC.A..GU.GA.....UGUGGCCUUUUU
CP002910        GAAAGACGCGCAUUUAUUAUCAuca..UCAUCCCUGA-AUCAGAGAUG..AAA.GUUU.GGCCAC.A..GU.GA.....UGUGGCCUUUUU
AM286415        GAAAGACGCGCAUUUGUUAUCA.....UCAUCCCUGUUAUCAGAGAUGUUAA...UUU.GGCCAC.A..GC.AA.....UGUGGCCUU-UU
CP002433        GAAAGACGCGCAUUUGUUAUCA.....UCAUCCCUGACAACAGAGAUGUUAA...UUC.GGCCAC.A..GU.GA.....UGUGGCCUU-UU
FP236842        GAAAGACGCGUAUUUGUUAUCAucaucUCAUCCCUGACAACAGAGAUGuUAA...UUUAGGCCAC.A..GU.GA.....CGUGGCCUUUUU
AP008232        GAAAGAUGCGCAUUUGUUAUCA.....UCAUCCCUGUUAACAGGAAUGUUAA...UUUA.GCCACAG...UuUC.....UGUGGCCUU-UU
CP001655        GAAAGACGCGCAUUUAUUAUCA.....UCAUCCCUAUUAUCAGAGAUGUU...UCUUUC.GCCAC.CcgGUAAcaaucgGGUGGCAUU-UU
#=GC SS_cons    ::::::::::::::::::::::.....:<<<<-<<<____>>>->>>>----.------<<<<<<.<..__.__.....>>>>>>>:::::
//
```

Figure 6: **MicA alignment containing additional homologs found using Plan B.** Two new sequences have been added to the alignment (*D. zeae*, CP001655; *S. glossinidius,* AP008232), adding structural and sequence diversity.

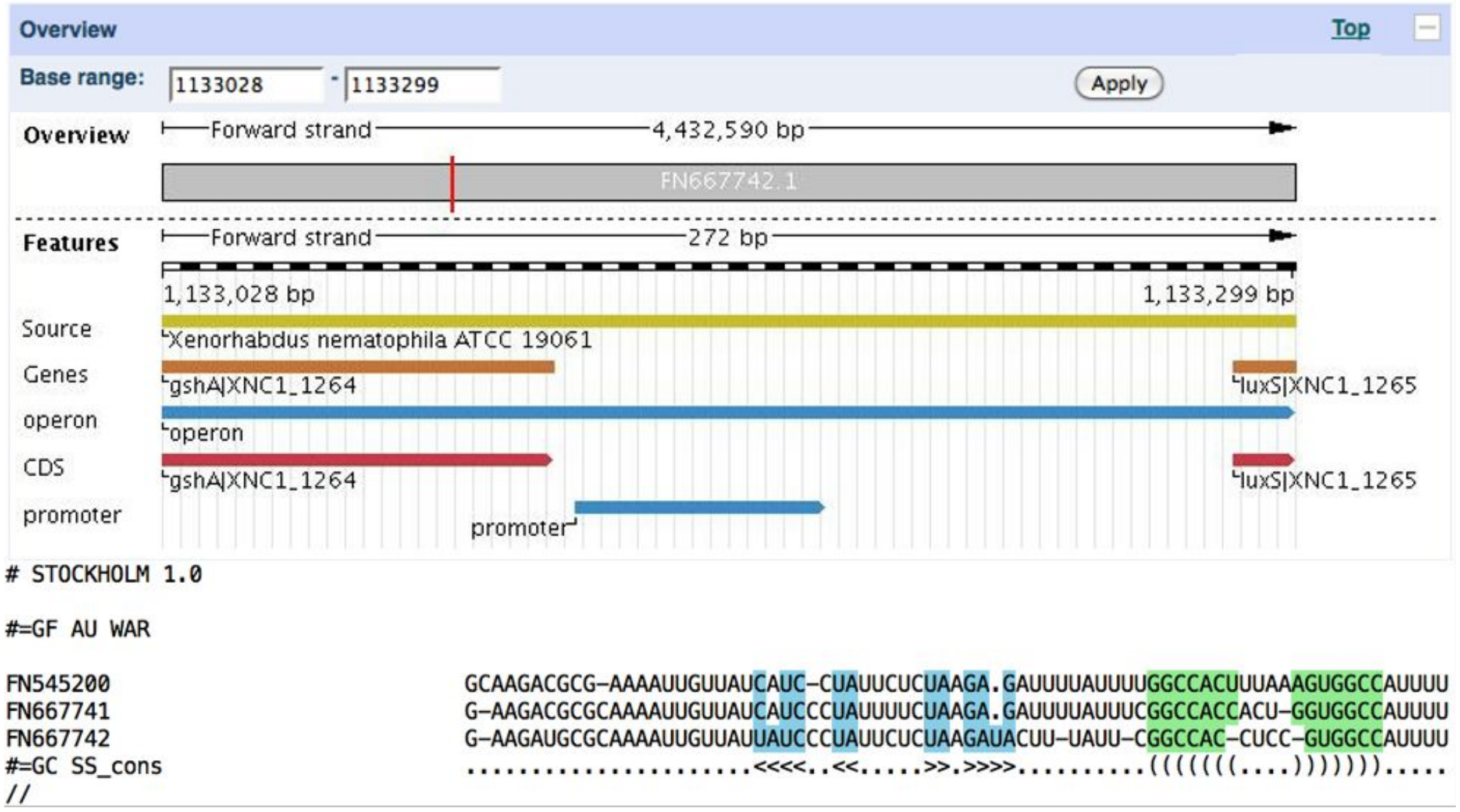

Figure 7: **Xenorhabdus MicA homologs found using Plan A.** The top panel shows the location of a putative *micA* homolog with a marginal Infernal *E*-value in *X. nematophila*, sharing synteny with established *micA* sequences in *E. coli*. The bottom panel shows a manually refined alignment of three putative *Xenorhabdus* MicA sequences, displaying structural and sequence similarity to our previously constructed alignment of enterobacterial MicA sequences.